\documentclass[11pt,a4paper]{article}
\pdfoutput=1

\usepackage[english]{babel}
\usepackage[utf8x]{inputenc}
\usepackage[T1]{fontenc}
\usepackage[a4paper,top=3cm,bottom=2cm,left=3cm,right=2.8cm,marginparwidth=1.75cm]{geometry}
\usepackage{amsmath}
\usepackage{graphicx}
\usepackage[colorinlistoftodos]{todonotes}
\usepackage[colorlinks=true, allcolors=blue]{hyperref}
\usepackage{graphicx}
\usepackage{subcaption}
\usepackage{authblk}
\usepackage[square,numbers]{natbib}

\title{\textbf{A measurement of absolute efficiency of the ARAPUCA photon detector in Liquid Argon}}
\author[a]{Dante Totani}
\author[b]{Gustavo Cancelo}
\author[b]{Flavio Cavanna}
\author[b]{Carlos O. Escobar}
\author[c]{Ernesto Kemp}
\author[d]{Franciole Marinho}
\author[e]{Laura Paulucci}
\author[f]{Dung D. Phan}
\author[g]{Stuart Mufson}
\author[g]{Chris Macias}
\author[h]{David Warner}
\affil[a]{Universit\`{a} degli Studi dell'Aquila\\L'Aquila, 67100, Italia}
\affil[b]{Fermi National Accelerator Laboratory\\Batavia, IL 60510, USA}
\affil[c]{Universidade Estadual de Campinas\\Campinas - SP, 13083-970, Brazil}
\affil[d]{Universidade Federal de S\~{a}o Carlos\\Araras - SP, 13604-900, Brazil}
\affil[e]{Universidade Federal do ABC\\Santo Andr\'{e} - SP, 09210-580, Brazil}
\affil[f]{University of Texas at Austin, Austin, TX 78712, USA}
\affil[g]{Indiana University\\Bloomington, IN 47405, USA}
\affil[h]{Colorado State University\\Fort Collins, CO 80523, USA}

\begin{document}
\maketitle

\begin{abstract}
In the Fall of 2017, two photon detector designs for the Deep Underground Neutrino Experiment (DUNE) Far Detector were installed and tested in the TallBo liquid argon (LAr) cryostat at the Proton Assembly (PAB) facility, Fermilab. The designs include two light bars developed at Indiana University and a photon detector based on the ARAPUCA light trap engineered by Colorado State University and Fermilab. The performance of these devices is determined by analyzing 8 weeks of cosmic ray data. The current paper focuses solely on the ARAPUCA device as the performance of the light bars will be reported separately. The paper briefly describes the ARAPUCA concept, the TallBo setup, and focuses on data analysis and results.
\end{abstract}

\section{Introduction}
The efficiency of photon detectors is of paramount importance for large volume LAr experiments. The detection of scintillation light generated as charged particles traverse a large liquid argon time-projection chamber (LArTPC) adds valuable information to the study of weakly-interacting particles. Most importantly, the leading edge of the scintillation light pulse yields sub-mm precision in reconstructing the absolute position of the event in the drift direction~\cite{Baller_2014}. In addition, the scintillation light can provide the trigger for baryon number violation events, such as proton decay and neutron-antineutron oscillations, and supernova neutrinos, as well as improve rejection of uncorrelated cosmic backgrounds. Given the enormous volume of future experiments such as DUNE (Deep Underground Neutrino Experiment)~\cite{2018arXiv180710334D}, the photon detectors must cover a large surface area in a cost-effective manner. Several technologies and implementations have been proposed and tested. A relatively novel technology is based on a light trapper device designed by A. Machado and E. Segretto named ARAPUCA~\cite{Machado:2016jqe}. ARAPUCAs are able to increase the effective photon collection area while keeping the sensor area small. The latter property allows a relatively high photon collection efficiency at a reasonable cost. The following paper reports on the design and evaluation of a photon detector unit composed of eight ARAPUCA cells. The detector was placed along a photon detector plane that also included two wavelength shifting light guides from the photon detector group at Indiana University~\cite{IU_paper}. The combined photon detector set operated for eight weeks in the TallBo LAr cryostat at the PAB facility, Fermilab.
\begin{figure}[t]
\begin{subfigure}{0.4\textwidth}
\includegraphics[width=\textwidth]{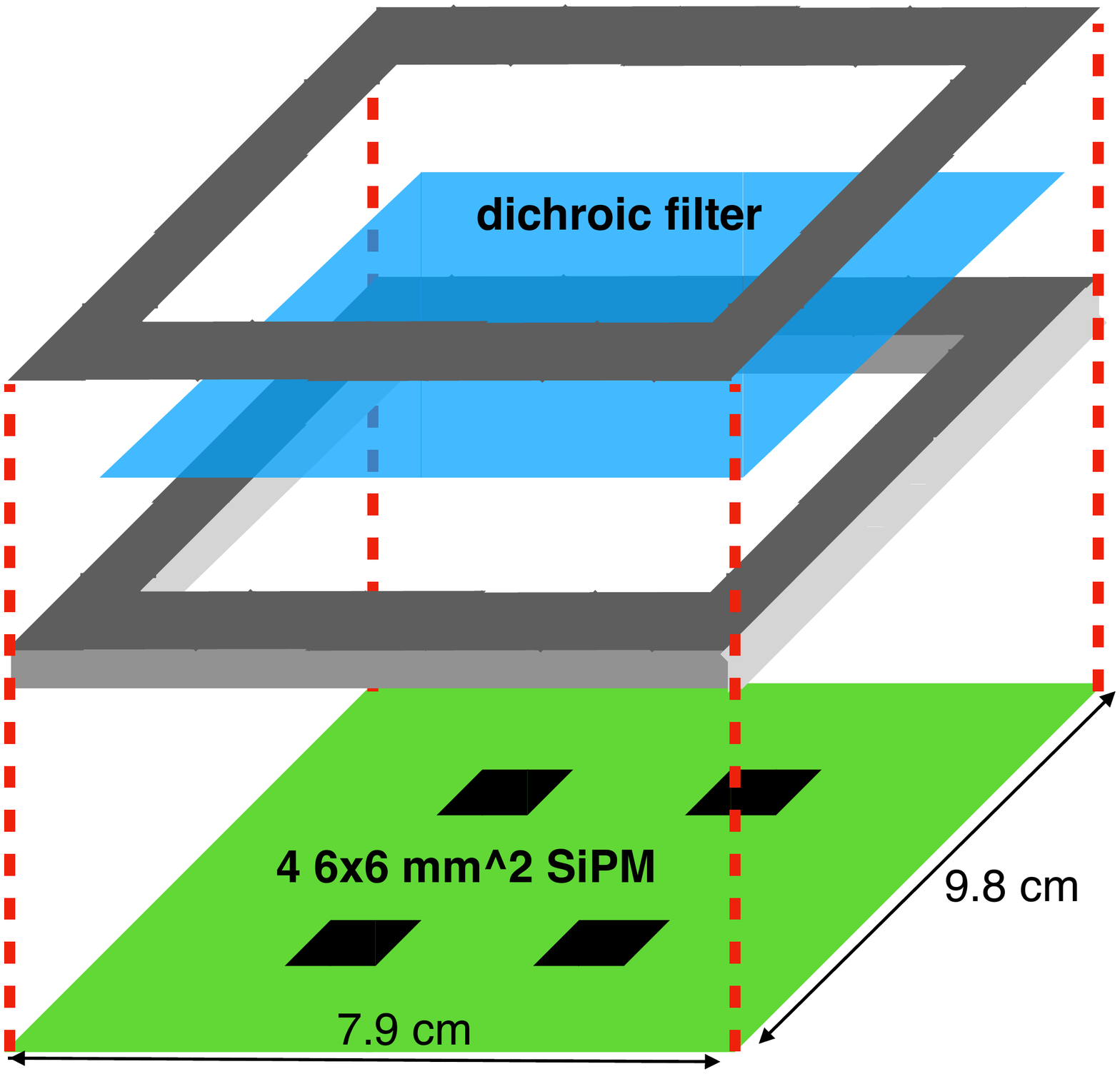}
\caption{ARAPUCA mechanical design}
\label{fig:ARAPUCA_b}
\end{subfigure}
\begin{subfigure}{0.6\textwidth}
\includegraphics[width=\textwidth]{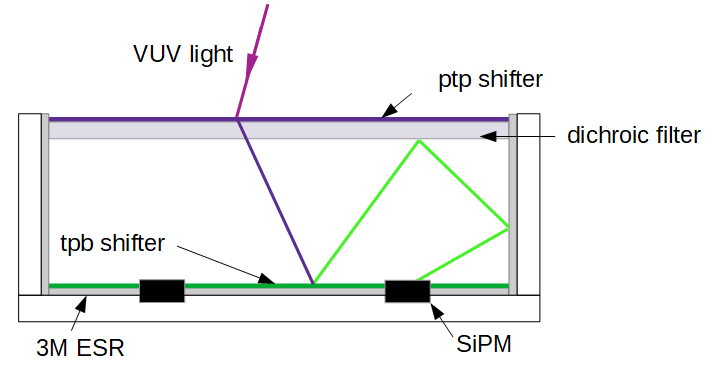}
\caption{ARAPUCA conceptual design}
\label{fig:ARAPUCA_a}
\end{subfigure}
\caption{ARAPUCA light trapper device.}
\label{fig:wl1}
\end{figure}

\begin{figure}[b]
\begin{subfigure}{0.5\textwidth}
\setlength{\captionmargin}{0.07\textwidth}
\includegraphics[width=\textwidth]{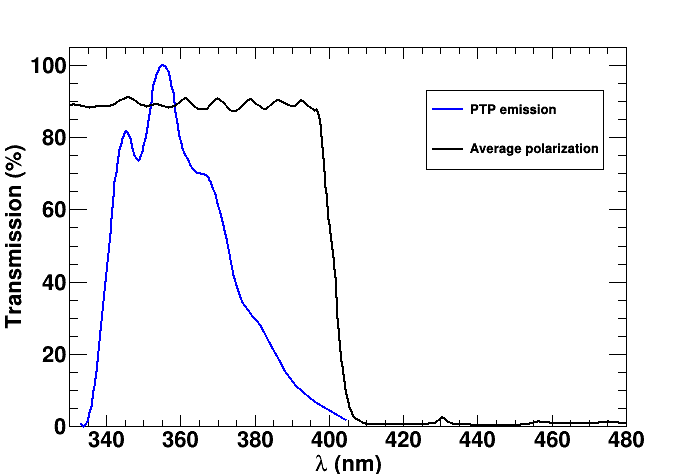}
\caption{Dichroic filter transmission and \mbox{p-terphenyl} emission spectra}
\label{fig:arapuca1stWLS}
\end{subfigure}
\begin{subfigure}{0.5\textwidth}
\setlength{\captionmargin}{0.07\textwidth}
\includegraphics[width=\textwidth]{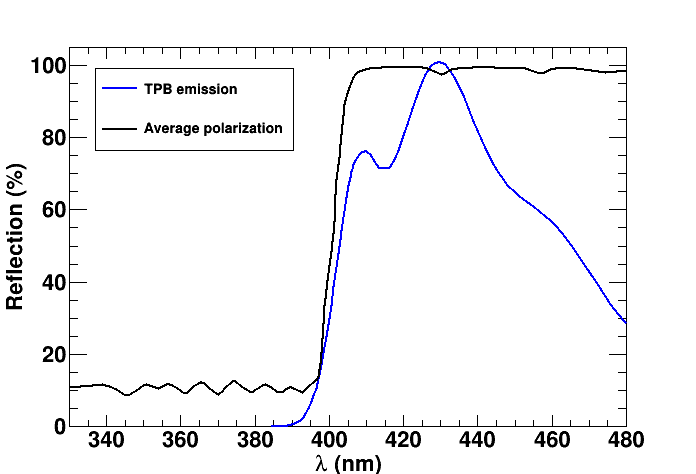}
\caption{Dichroic filter reflection and \mbox{TPB emission} spectra}
\label{fig:arapuca2ndWLS}
\end{subfigure}
\caption{Double coated dichroic filter and wavelength shifter spectra}
\label{fig:wl}
\end{figure}

\subsection{The ARAPUCA Light Trap}
The $128\, nm$ scintillation light from interactions of charged particles in LAr is not detectable directly by affordable sensors such as the traditional photomultipliers (PMTs) or the newer silicon photomultipliers (SIPMs) without appropriate wavelength shifter coating. Among available photon detectors, SIPMs are gaining popularity due to its superior quantum efficiency (QE), desired physical and electrical properties as well as a low cost of production. However, due to their small size, their effective light collection area per unit is easily surpassed by PMTs. The ARAPUCA concept~\cite{Machado:2016jqe} brings up a solution to this problem. As shown in Figure~\ref{fig:ARAPUCA_a} the ARAPUCA consists of two wavelength shifters, a dichroic filter, a highly reflective box and photosensors (SIPMs). The face of the box is used to augment the photon collection area. The VUV photons that enter the light collecting surface of an ARAPUCA device are wavelength-shifted by $200\, \mu g/cm^2$ of p-terphenyl deposited on the external face of a dichroic filter. The dichroic filter allows the converted photons to go through and enter the box. Those photons are wavelength shifted a second time using $250\, \mu g/cm^2$ of tetra-phenyl butadiene (TPB) deposited on the internal face of the dichroic filter. Since the wavelength cutoff of the dichroic filter is in between the emission spectrum of p-terphenyl and the emission spectrum of TPB, the twice shifted photons remain trapped inside the reflective box and bounce off the walls until they hit the sensors. Figure~\ref{fig:ARAPUCA_b} shows one of the mechanical designs of the ARAPUCA. Several iterations of ARAPUCAs have been studied with variable box sizes, number of SIPMs and location of the SIPMs. In one of the experiments with previous versions of the ARAPUCAs, also performed at TallBo during the Spring 2017 with a radioactive source to excite the LAr, we measured an \mbox{efficiency $\sim 0.4 \%$~\cite{ArapucaLIDINE}}, considerably lower than the values reported with the new configuration used in this work (see section~\ref{eficana}). The Fall 2017 TallBo experiment used 8 ARAPUCA units of $9.8\,cm \times 7.9\,cm$ with 4 SensL $6\,mm \times 6\,mm$ SIPM biased at $25.5\, V$. Figure~\ref{fig:arapuca1stWLS} and~\ref{fig:arapuca2ndWLS} show that the emission spectra of p-terphenyl and TPB do not overlap; and the wavelength cutoff of the dichroic filter is located between the two spectra. The peak of the p-terphenyl spectrum is $360\, nm$, the peak of the TPB spectrum is $430\, nm$, and the cutoff of the filter is $400\, nm$.

\begin{figure}[h]
\centering
\includegraphics[width=0.5\textwidth]{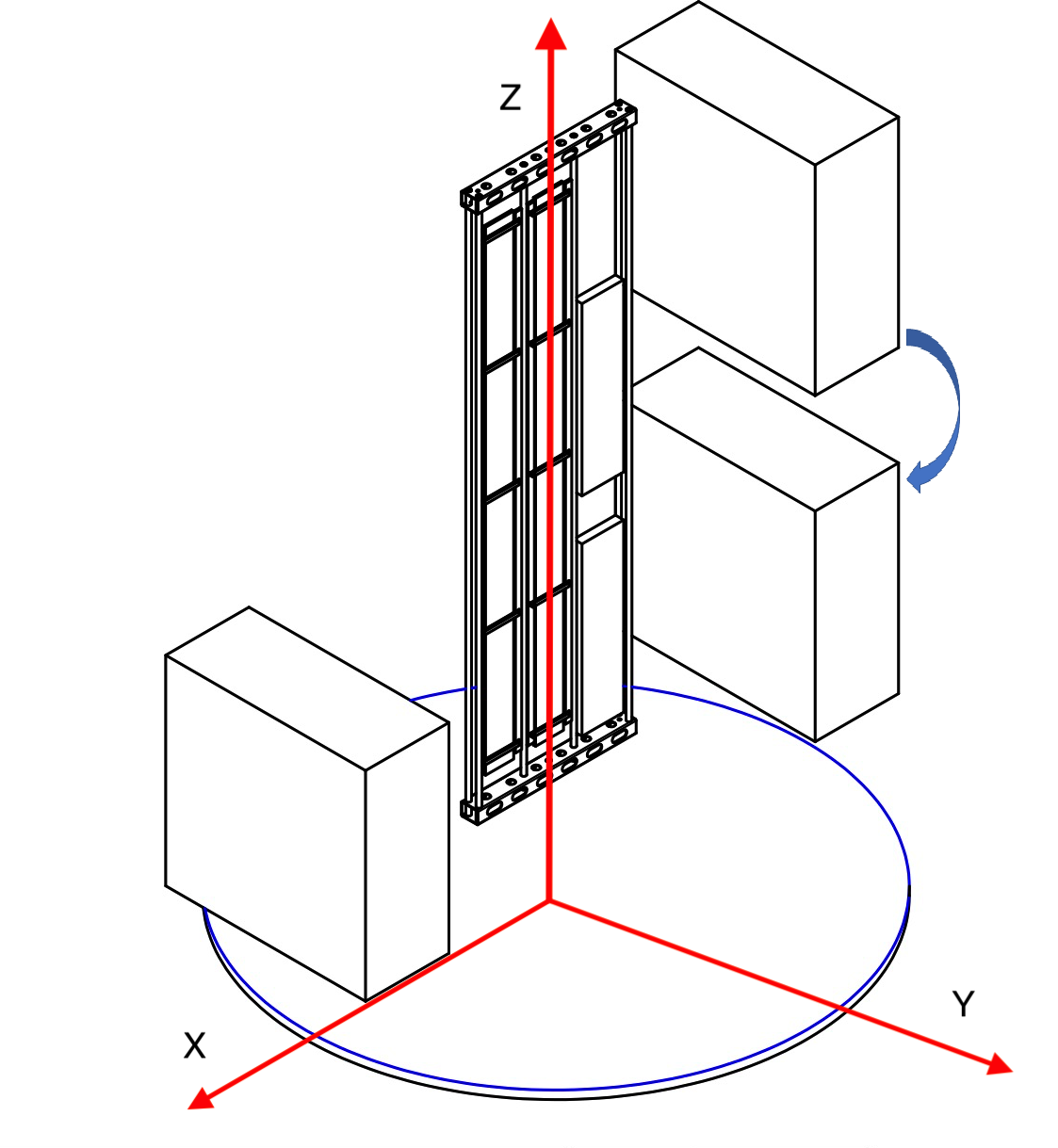}
\caption{3D view of the TallBo LAr cryostat.}
\label{fig:tallbo7}
\end{figure}

\section{Experimental Setup}

\subsection{The TallBo cryostat and hodoscope}

\begin{figure}
\begin{subfigure}{\textwidth}
\includegraphics[width=\textwidth]{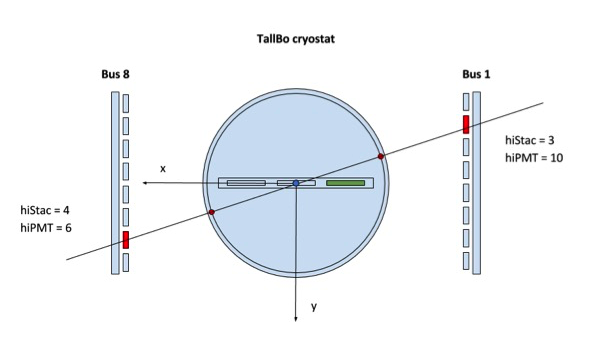}
\caption{Top view.}
\label{fig:topview}
\end{subfigure}
\begin{subfigure}{\textwidth}
\includegraphics[width=\textwidth]{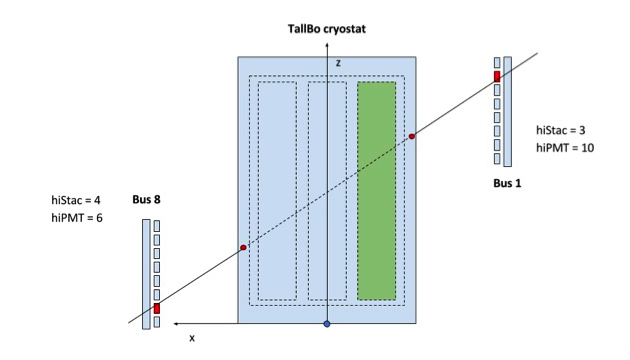}
\caption{Front view.}
\label{fig:frontview}
\end{subfigure}
\caption{Cross-sectional views of the TallBo LAr cryostat.}
\end{figure}

The vessel’s external diameter is $61.82\, cm$ and the inner diameter is $55.88 \, cm$. The full capacity of TallBo is $460\, l$. For this experiment TallBo was filled with approximately $350\, l$ of liquid argon. The total thickness of stainless steel presented to a cosmic ray particle that crosses the cryostat is $5.08 \, cm$. The liquid Argon purity in TallBo was monitored using commercial gas analyzers. During the period of the data acquisition the levels of $N_2$, $H_2 O$, and $O_2$ were well below $1\,ppm$ ($ppm=$ part per milion). Tipical values were: $N_2 \sim 230\,ppb$, $H_2 O\sim 2-3 \,ppb$ and $0_2\sim 30 \, ppb$ ($ppb=$ part per bilion).\\
The experiment's trigger used a set of two scintillation paddles and a tracking mechanism based on scintillation paddles and hodoscopes. The hodoscopes were used before in the CREST baloon flight experiment~\cite{Coutu:2011zz, blanchpaper}. The hodoscope modules were installed on opposite sides of the TallBo cryostat to select single-track cosmic-ray muons passing through the LAr volume~\cite{blanchpaper}. Figure~\ref{fig:tallbo7} shows a 3D view of the photon detector plane inside the cryostat and the two hodoscope blocks (in white), one of which was moved from a high position to a low position to trigger on high angle and low angle minimum ionizing particles (MIPs) respectively. Each hodoscope module consists of 64 2-inch diameter barium-fluoride crystals, coated with TPB. Each crystal is monitored by a 2-inch PMT. The crystals are arranged as $8\,\times\, 8$ matrix. Since the hodoscope matrix elements are very sensitive to extraneous photon activity and have a high dark count rate, to remove extraneous events two scintillator panels covering the entire hodoscope face were placed between each hodoscope module and the TallBo dwear. These panels are individually read out by PMTs. The readout system was then triggered by four-fold coincidence logic that required at least one hit in both hodoscope modules as well as one hit in their adjacent scintillator planes in a coincidence window of 150 ns. Events were further filtered offline by requiring one and only one hit in each hodoscope module to reject showers. Together triggering on cosmic rays rejecting showers the requirement of having a single crystal fired per hodosocpe module gave the geometric position of the tracks. Figure~\ref{fig:topview}~and~\ref{fig:frontview} show, from the top and from the front of the cryostat, how a track is identified and located by the hodoscope. 
\begin{figure}
\centering
\includegraphics[width=0.8\textwidth]{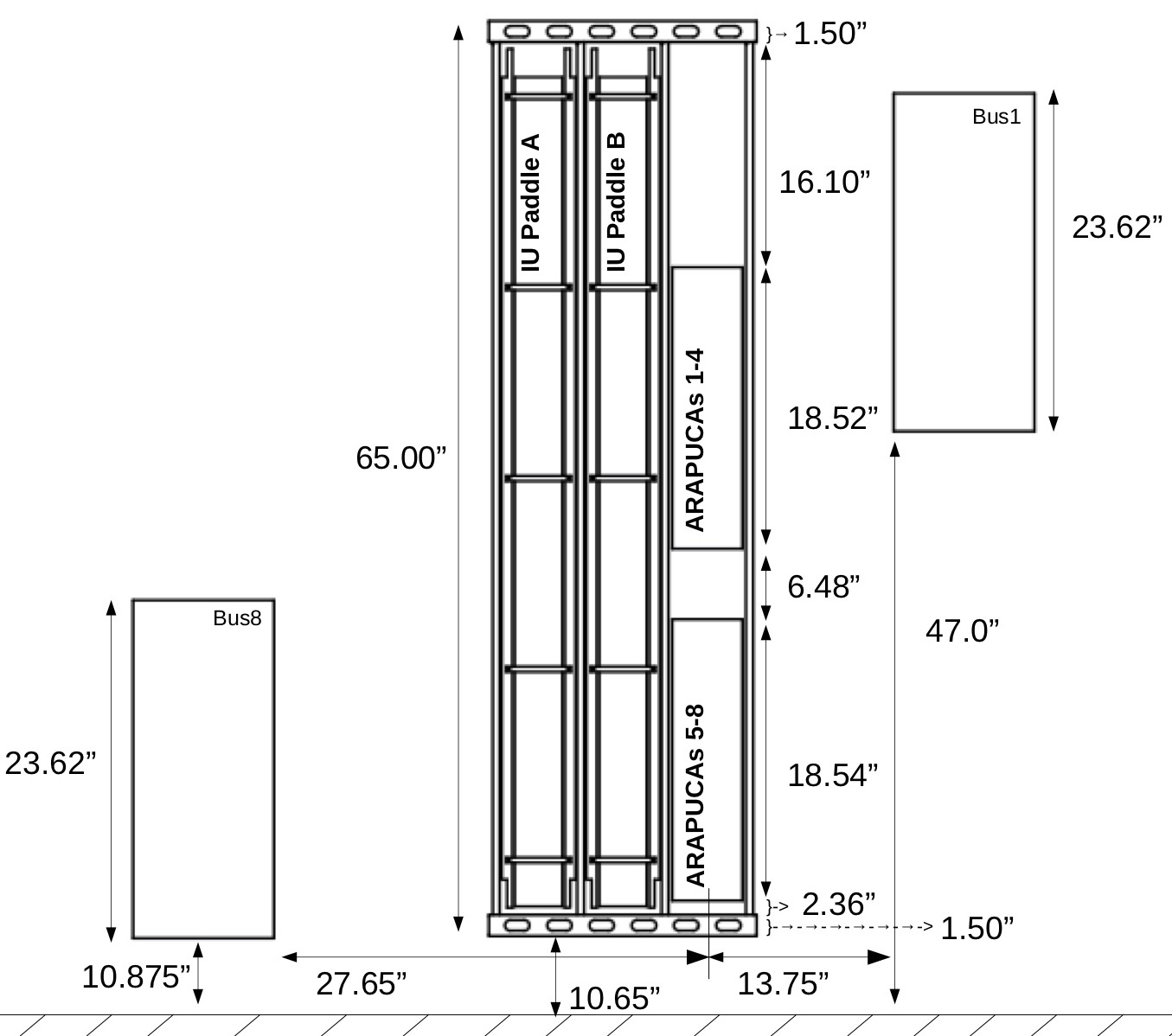}
\caption{Detector plane configuration.}
\label{fig:hilow}
\end{figure}

\subsection{The ARAPUCA array}
As shown in Figure~\ref{fig:hilow}, the 8 ARAPUCAs are divided into two frames of 4 ARAPUCAs each and occupy one third of the photon detector plane. They were located in a side position with respect to the vertical central axis plane. Each ARAPUCA has dimensions of $12\,cm \,\times \,9.5\,cm \, \times \,1\,cm$. The filter windows are $10 \,cm \,\times\, 8 \, cm$. Two sets of four contiguous ARAPUCAs are placed as shown separated by a $14.5\, cm$ gap. The values in Figure~\ref{fig:hilow} are given in inches.

\subsection{Data Acquisition System}
Signals are processed by the SiPM Signal Processor (SSP) module designed by the Electronics Group of the High Energy Physics division at Argonne National Laboratory~\cite{SSP}. The SSP is a $14\,bit$, $150\,MS/s$, 12-channel waveform digitizer DAQ. The ADC has a full-scale dynamic range of $2\,V$ and a preamplifier gain of $18.8 \, V/V$. For typical SiPM gains and ganging configurations the SSP allows large signals equivalent $\sim 1000$ PEs before the ADC saturates. The SSP input noise is $\sim 16\,\mu V$. Each acquired waveform contains 1950 ADC values sampled at $150\, MS/s$, aggregating to an acquisition time of $13 \, \mu s$ Each sample bin, $6.67\, ns$ long, is called $tick$. The SSP can trigger internally on each individual channel or externally via external trigger input. The self-trigger mode was used to take calibration data while the external trigger mode with the hodoscope as the trigger source was employed for the cosmic ray run.

\subsection{Trigger issues}\label{trigger_issues}
 The hodoscope trigger malfunctioned during the experiment. A mistake in the way the coincidence logic was set up caused the trigger to fire more often than expected. The logic performed an OR instead of an AND in two of the channels. Therefore, the trigger not only fired on valid tracks but also on coincidental PMT dark counts of the hodoscope arrays located at the left and right of the dewar during the defined trigger window ($150\,ns$). The collected data set contained a large amount of unusable data that did not represent valid tracks. As shown in the next sections, most of the events were triggered by coincidence of PMT dark counts and contained no signals from the photon detectors, neither in the ARAPUCAs nor the IU light guides. Other spurious events contained partial tracks that did not generate a trigger but were captured by chance during a dark count coincidence event. The hodoscope trigger problem made the analysis of the data challenging, since offline filters had to be devised and applied to the data. However, it also showed the potential of the ARAPUCA architecture given that those filters could have not been applied without the segmentation property of the ARAPUCAs, as will become clear in the next sections of the paper.

\section{Data analysis}

\subsection{Calibration}

\begin{figure}[h]
\begin{subfigure}[b]{0.5\textwidth}
\includegraphics[width=\textwidth]{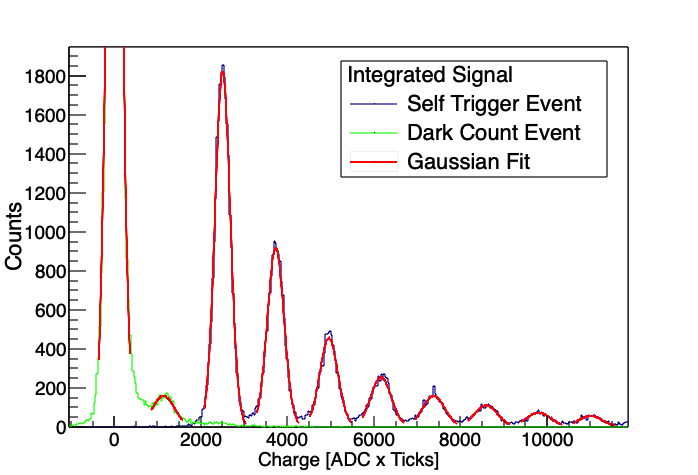}
\setlength{\captionmargin}{0.07\textwidth}
\caption{Self triggered and dark count data spectra. In red is shown the multi gaussian fit.}
\label{fig:specal}
\end{subfigure}
\begin{subfigure}[b]{0.5\textwidth}
\includegraphics[width=\textwidth]{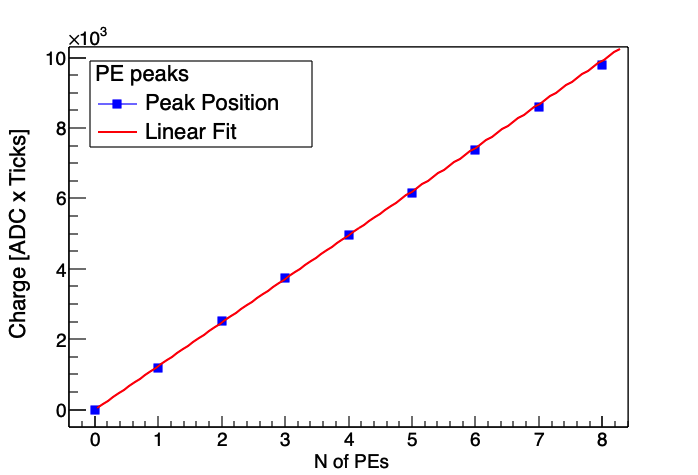}
\setlength{\captionmargin}{0.07\textwidth}
\caption{Linear fit for the peak positions. Statistical error bars are smaller than symbols.}
\label{fig:pecal}
\end{subfigure}
\begin{subfigure}[b]{0.5\textwidth}
\includegraphics[width=\textwidth]{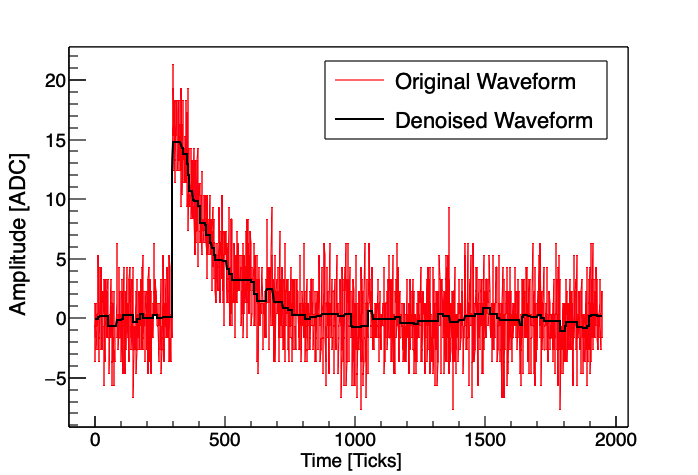}
\setlength{\captionmargin}{0.07\textwidth}
\caption{Original and denoised waveform \mbox{comparison}.}
\label{fig:calcheck1}
\end{subfigure}
\begin{subfigure}[b]{0.5\textwidth}
\includegraphics[width=\textwidth]{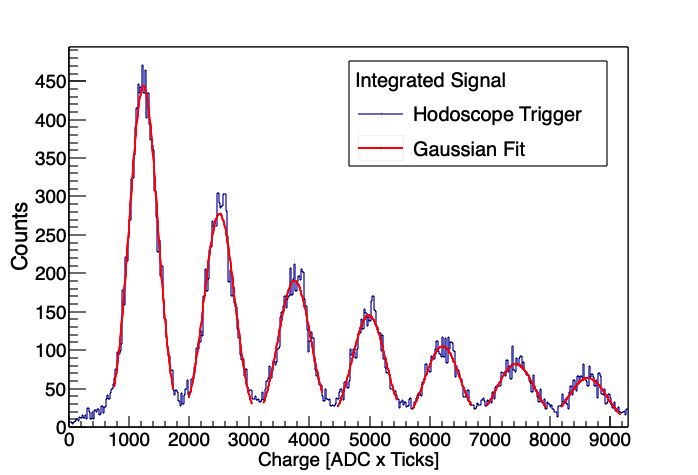}
\setlength{\captionmargin}{0.07\textwidth}
\caption{Hodoscope triggered data spectrum and multi gaussian fit.}
\label{fig:calcheck2}
\end{subfigure}
\caption{Calibration procedures for ARAPUCA 1. The slope of linear fit (b) gives the calibration in $(ADC\cdot Ticks)/PE$.}
\end{figure}

During calibration runs the ARAPUCA signals were acquired by the SSP single channel self trigger mode, using an amplitude threshold slightly above 1~PE. Since the threshold was higher than the single photon electron the histogram is able to exhibit the $2^{nd}$ peak and higher \mbox{(Figure~\ref{fig:specal} blue histogram)}. The first PE was recovered looking at dark counts in a different integration window \mbox{(Figure~\ref{fig:specal} green histogram)}, far from the trigger point. The first integration window, containing the self trigger point, was in the first half of the waveform (from 100 to 900 $ticks$) and the second window, containing possible dark counts, in the second half of the waveform (from 900 to 1700 $ticks$). Figure~\ref{fig:pecal} shows the linearity of the calibration. The linear fit in Figure~\ref{fig:pecal} is obtained using the means of the Gaussian distribution peaks of Figure~\ref{fig:specal}. The integrated charge value ($ADC\cdot Ticks$) due to one PE can be obtained by the slope of the straight line.\\
A second procedure was developed to cross check values using the calibration constants obtained from the calibration run. A smaller set of data triggered with the hodoscope was used. Baseline and noise levels were obtained as the mean and root mean square (RMS) for the first 100 points of the signal amplitude for each channel per event. After baseline subtraction a filtering algorithm was used that suppresses random noise but keeps the main features of the signal, in particular the fast transitions~\cite{denoise}. A peak finding algorithm was then devised to find peaks and integrate the charge under a peak. Figure~\ref{fig:calcheck1} shows an example of a peak found. The red line shows the raw data and the black line indicates the filtered signal. The threshold for detecting a pulse has been set to $3\sigma$ of the baseline noise (e.g. 8.2 ADC). Figure~\ref{fig:calcheck2} shows the spectrum obtained with peaks from ARAPUCA 1 (the same used to show the self trigger events analysis). Notice that the pedestal is almost suppressed by the denoising procedure and threshold requirement. A multi Gaussian fit was performed to obtain the charge value for each peak. A linear fit of the peaks as a function of the photon electron number provided the average calibration constants which are in  agreement with the self trigger events analysis.

\subsection{Data analysis: background rejection}\label{bgr}

\begin{figure}[h]
\centering
\includegraphics[width=0.9\textwidth]{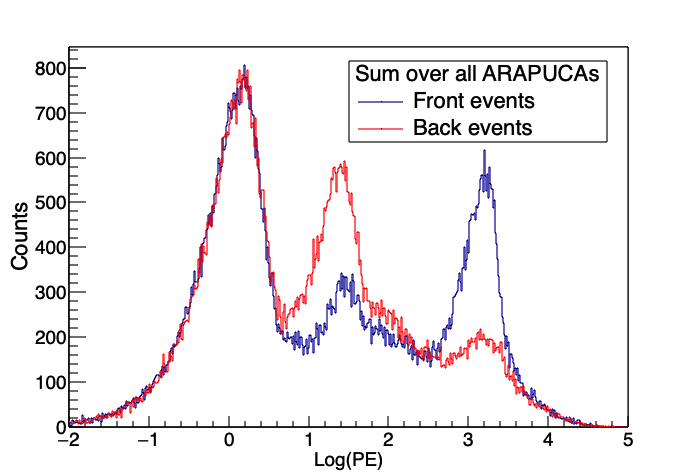}
\caption{Log(PE) spectra for events triggered as "front" (blue) and "back" (red). The same number of events for both spectra are used, this normalization is reflected in having the same amount of empty events (first peak in both spectra). The excess of events in the second peak of the red spectrum respect the blue one is interpreted as back tracks associated to the right trigger information. In the same way the excess of events in the third peak of the blue spectrum respect the red one is interpreted as front tracks associated to the right trigger information. All common parts of the spectra are interpreted as background: empty events (first peak) and spurious hodoscope coincidences.}
\label{fig:FBspectra}
\end{figure}

\begin{figure}[h]
\begin{subfigure}{0.5\textwidth}
\captionsetup{width=.9\textwidth}
\includegraphics[width=\textwidth]{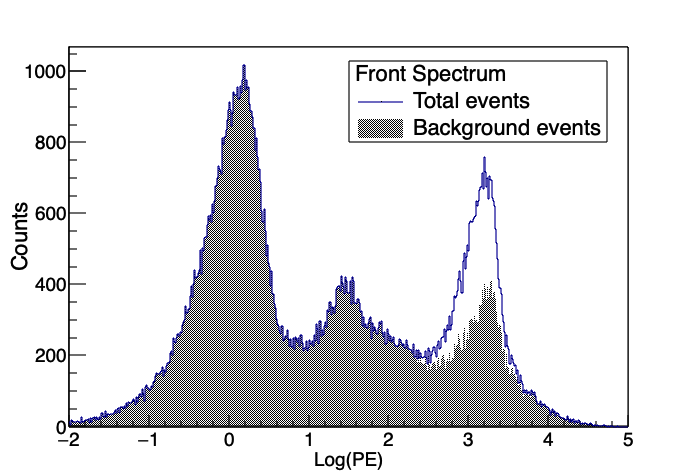}
\caption{Background.}
\label{fig:bkss}
\end{subfigure}
\begin{subfigure}{0.5\textwidth}
\captionsetup{width=.9\textwidth}
\includegraphics[width=\textwidth]{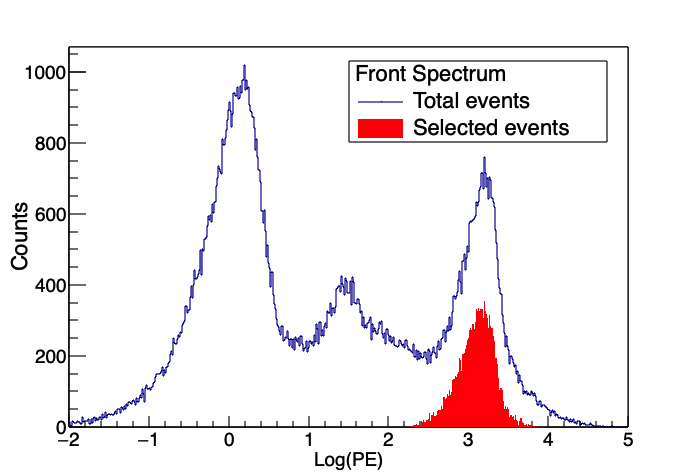}
\caption{Selected events.}
\label{fig:bfgv}
\end{subfigure}
\caption{Background and selected events obtained through cuts applied on the number of PEs collected by each ARAPUCA, got comparing front and back spectra.}
\label{fig:spectra123zl}
\end{figure}

\begin{figure}
\begin{subfigure}[b]{0.5\textwidth}
\includegraphics[width=\textwidth]{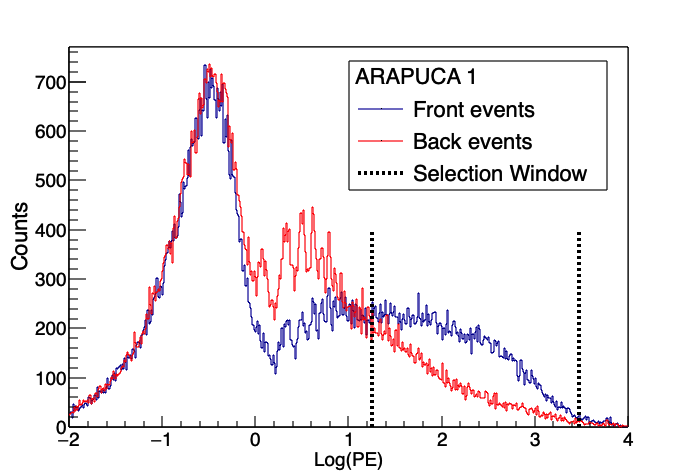}
\end{subfigure}
\begin{subfigure}[b]{0.5\textwidth}
\includegraphics[width=\textwidth]{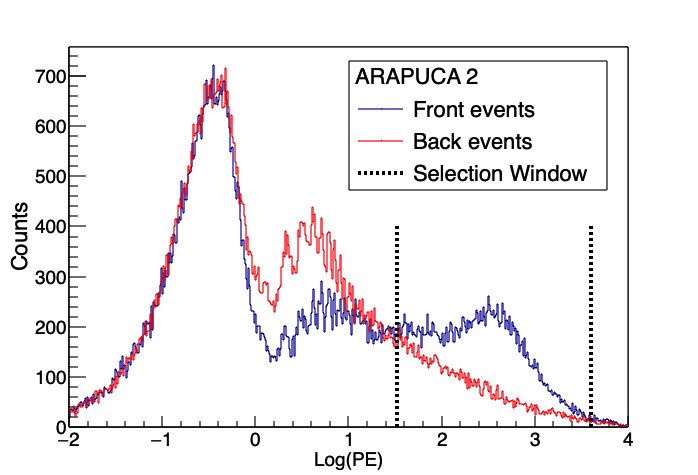}
\end{subfigure}
\begin{subfigure}[b]{0.5\textwidth}
\includegraphics[width=\textwidth]{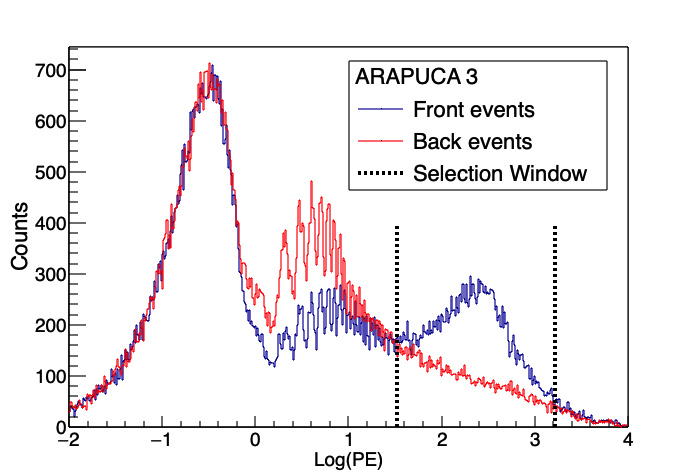}
\end{subfigure}
\begin{subfigure}[b]{0.5\textwidth}
\includegraphics[width=\textwidth]{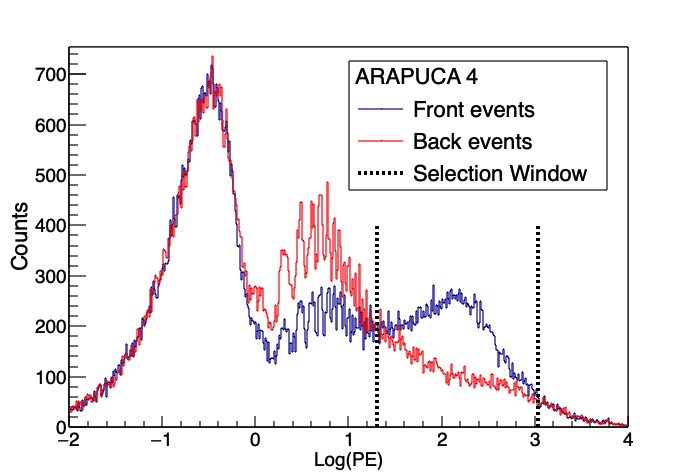}
\end{subfigure}
\begin{subfigure}[b]{0.5\textwidth}
\includegraphics[width=\textwidth]{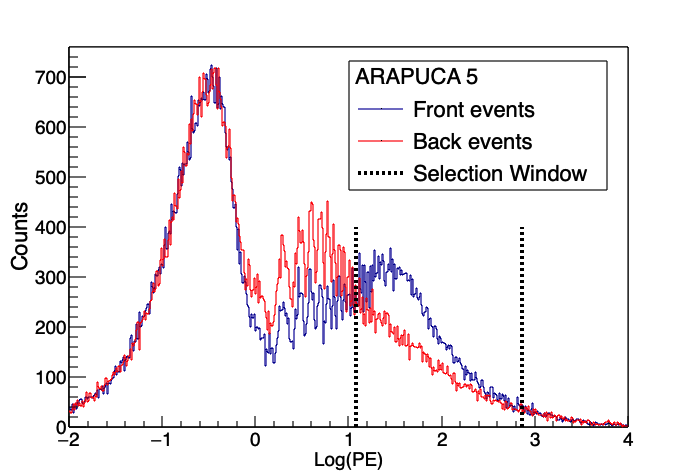}
\end{subfigure}
\begin{subfigure}[b]{0.5\textwidth}
\includegraphics[width=\textwidth]{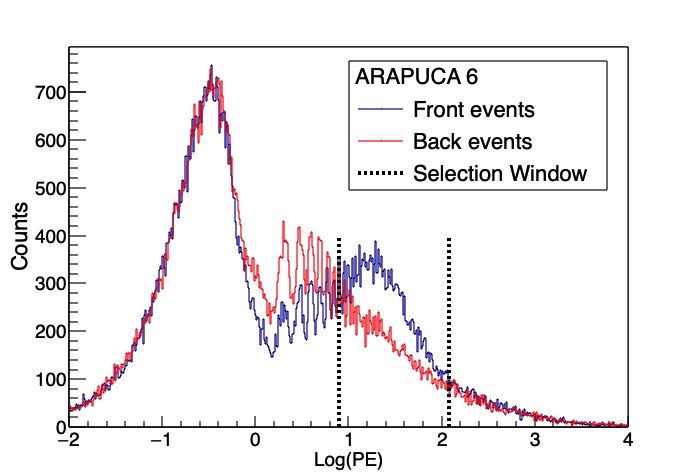}
\end{subfigure}
\begin{subfigure}[b]{0.5\textwidth}
\includegraphics[width=\textwidth]{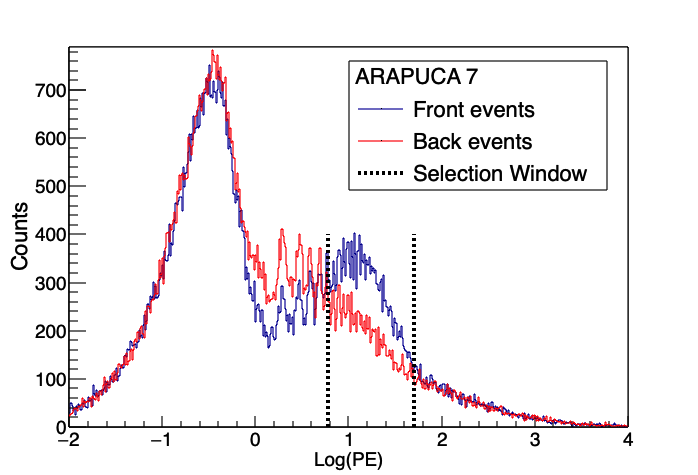}
\end{subfigure}
\begin{subfigure}[b]{0.5\textwidth}
\includegraphics[width=\textwidth]{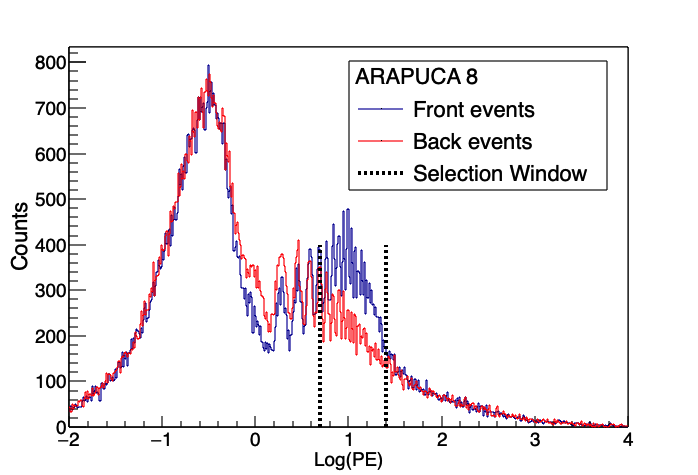}
\end{subfigure}
\caption{Front (blue) and back (red) Log(PE) spectra for the eight ARAPUCAs, and the cut applied (black). The cut consists in removing events for which the back spectrum (red) has more events respect to the front spectrum (blue).}
\label{fig:ar1to8}
\end{figure}

As mentioned in subsection \ref{trigger_issues} most of the data recorded by the experiment was due to the background, since the majority of the triggers were generated by coincidental dark counts in the hodoscope PMTs. Nevertheless, the system has also properly triggered on real tracks (i.e. MIPs), although these events only constituted a small fraction of the stored data, of the order of $5\%$ to $10\%$.\\
Since the dark counts are uncorrelated with light in the cryostat most events of that type were empty or captured incidental background light. Most of these events collected few PEs from radiogenics or very low light background in the TallBo cryostat. A fraction of those events also captured an incidental track that fell within the $150\,ns$ trigger window. Most of those tracks are partial tracks that only passed through one hodoscope arm or none at all. In either case the track position information given by the hodoscope is incorrect, since it corresponds to the hot PMTs which generated the trigger and not to the passage of the track.\\
The tested version of ARAPUCAs were single sided detectors having filters on only one side. Therefore only tracks passing in front of the ARAPUCA plane could  be seen. In this particular occasion, due to the high trigger rate, the comparison between the spectra of events passing in front of the single sided ARAPUCAs and behind them was useful to do event discrimination. It was possible to get information comparing the spectra of events triggered as tracks located entirely in front of the plane containing the windows and behind it. That allowed tracks uncorrelated with trigger position information to be filtered out.\\
 \\
As a first step we considered the spectrum of the number of PEs seen by all \mbox{ARAPUCAs} in the detector, for a given event ($PE_{tot}$). We obtained individual spectra for two groups of events. Group~1 (blue spectrum of Figures~\ref{fig:FBspectra}) has all events for which the trigger geometry given by the hodoscope labeled a track as passing in front of the ARAPUCA window plane (front events) and Group~2 (red spectrum of Figures~\ref{fig:FBspectra}) with all events passing behind the ARAPUCA window plane (back events). Events that have an entry point in the front and the other one in the back or vice-versa were discarded, like the example in Figures~\ref{fig:topview}. The difference between the back and front spectra is the only feature which can be associated to the information coming from the hodoscopes. To enhance the peak display in the spectra, instead to use the PE number, it has been chosen to show the common logarithm of the PE number in plots.\\
 \\
The spectra, reported in Figures~\ref{fig:FBspectra}, show three peaks: 
\begin{itemize}
\item the first, the largest one, is composed of events for which two crystals fired but negligible light is seen by any ARAPUCA, indicating these are completely random triggers, due to the high hodoscope rates. The fact that both back and front spectra show the first peak identically confirms its origin. The same number of events for both spectra are used. 
\item the excess of events in the second peak of the red spectrum respect the blue one is interpreted as back tracks associated with a MIP triggering the crystals for a track located entirely behind the ARAPUCAs plan (back track). 
\item in an equivalent way, the excess of events in the third peak of the blue spectrum respect the red one is interpreted as front tracks associated with a MIP triggering the crystals for a track located entirely in front of the ARAPUCAs plan (front track).  
\end{itemize}
All common parts of the spectra are interpreted as background: empty events (first peak) and spurious hodoscope coincidences.\\
 \\
Similar considerations can be made for each single ARAPUCA. Applying cuts on the minimum and maximum number of PEs collected by each ARAPUCA, it is possible to get a selected dataset of events. Single ARAPUCA cuts, shown as black lines in \ref{fig:ar1to8}, consist in removing events for which the back spectrum (red in \ref{fig:ar1to8}) has more events respect to the front spectrum (blue in \ref{fig:ar1to8}). The front and back spectra show a different behavior for each of the eight ARAPUCAs. The reason of that is the geometrical effect due to the position of the ARAPUCA cells respect the tracks selected by the hodoscope. For this reason the cut needed is different for each ARAPUCA, black lines in Figure~\ref{fig:ar1to8}.\\
 \\
Looking at the sum of the detected PE over the 8 ARAPUCA for the selected events, red spectrum in Figure~\ref{fig:bfgv}, they are compatible with the excess of events in the third peak of the blue spectrum in Figure~\ref{fig:FBspectra}. Events remained, black spectrum in Figure~\ref{fig:bkss}, are compatible with the common parts of the spectra in Figure~\ref{fig:bfgv}, interpreted as background.

\subsection{ARAPUCAs multi-channel information}

Each ARAPUCA is read out by an independent SSP channel, the segmented information can be used to study the track geometry. The number of PEs measured by each ARAPUCA must match the profile of the expected number of photons hitting the ARAPUCA array given a track geometry, with the track geometry provided by the hodoscope. The ratio between the number of PEs collected and the number of photons hitting each ARAPUCA depends only on their intrinsic features (geometrical dimensions, number of SiPM, wavelength shifter, dichroic filter). It should be constant and independent of the track geometry and the number of photons generated by a track in the LAr. Inconsistency of these ratio means we are observing a track not compatible with the geometry provided by the hodoscope. In that experiment the 8 cells of which the ARAPUCA array is composed, have the same features so we expect them to behave in the same way. The determination of the number of photons hitting each ARAPUCA is shown in the next section~(\ref{lp}).

\begin{figure}[h]
\centering
\includegraphics[width=0.5\textwidth]{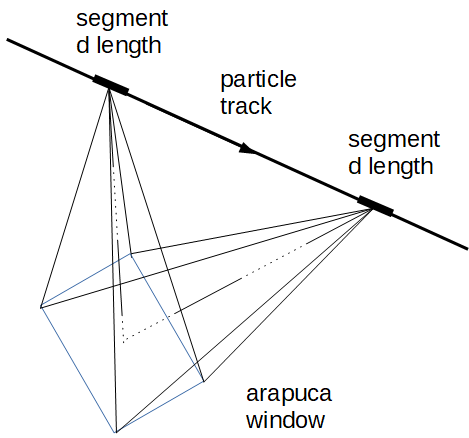}
\caption{Track angular acceptance.}
\label{trksegment}
\end{figure}

\subsection{Light pattern}\label{lp}
\begin{figure}[h]
\begin{subfigure}[b]{0.5\textwidth}
\includegraphics[width=\textwidth]{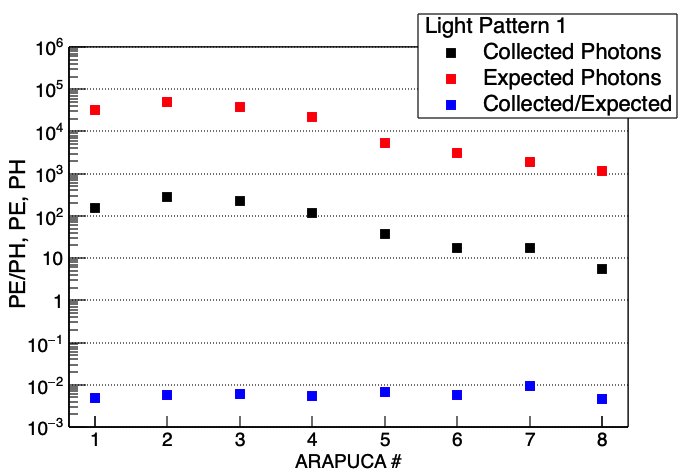}
        \caption{ 3}
        \label{fig:goodevent1}
\end{subfigure}
\begin{subfigure}[b]{0.5\textwidth}
\includegraphics[width=\textwidth]{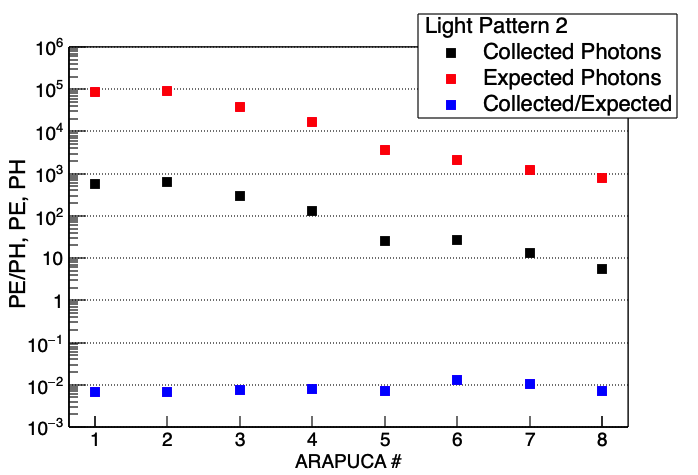}
        \caption{ }
        \label{fig:goodevent2}
\end{subfigure}
\begin{subfigure}[b]{0.5\textwidth}
\includegraphics[width=\textwidth]{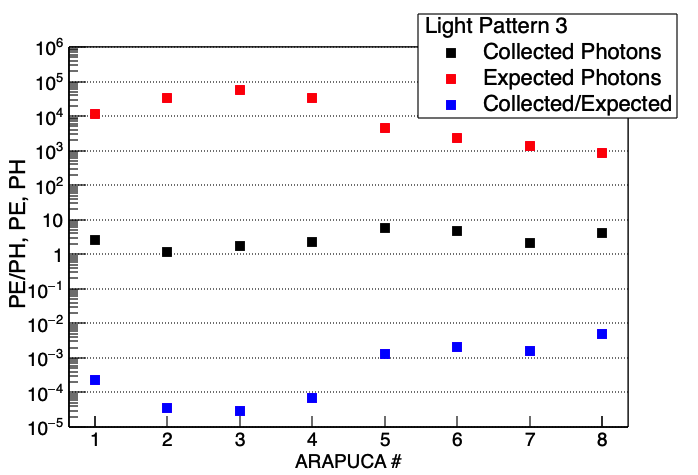}
        \caption{ }
        \label{fig:badevent1}
\end{subfigure}
\begin{subfigure}[b]{0.5\textwidth}
\includegraphics[width=\textwidth]{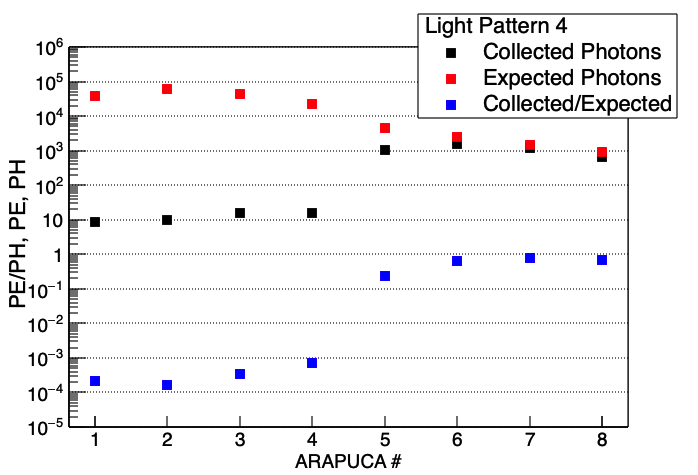}
        \caption{ }
        \label{fig:badevent2}
\end{subfigure}
\caption{Comparison between the amount of light expected to arrive in each ARAPUCA (red), the number of PEs extracted from the waveforms (black) and the ratio between them (blue). Figures (a) and (b) show events following the pattern expected, Figures (c) and (d) show events not correlated with the trigger information. Blue points show the ratio of measured to expected number of photons. Statistical error bars are smaller than symbols.}
\label{fig:pepattern}
\end{figure}
A charged particle crossing the volume of liquid argon will generate an amount of scintillation photons per unit of track length given by:
\begin{equation}
\frac{dN^{\gamma}}{dl} = \left< \frac{dE}{dx} \right> \rho Y_{\gamma} q_p,
\end{equation}
where $\left< \frac{dE}{dx} \right>$ is the specific deposited energy ($MeV cm^2/g$), $\rho$ is the density of the liquid in which the particle travels ($g/cm^3$), $Y_{\gamma}$ is the number of photons emitted per unit deposited energy in the medium ($Y_{\gamma} = 5.1\times 10^4\,\gamma/MeV$), and $q_p=0.78$ is the quenching factor~\cite{Doke:1988dp,2002JaJAP..41.1538D}. A crossing muon was assumed to be a MIP with a photon yield in liquid argon \mbox{$Y_{\gamma} q_p=4\times 10^4$ $photons/MeV$} and an energy deposit \mbox{$\left< \frac{dE}{dx} \right> \rho =2.12\, MeV/cm$}.\\
The number of photons that arrive at the detector window due to a muon passing the argon volume is estimated as the product of the track integrated angular acceptance, $A_{\Omega}$, and the number of emitted photons per unit length. This acceptance is defined numerically as:
\begin{equation}
A_{\Omega} = d\sum_{i=1}^{N}\Omega_i,
\end{equation}
where $d$ is the length of a small (i.e. differential) segment of the muon trajectory and $\Omega_i$ is the solid angle of the pyramid with its apex at the center of that segment and its base given by the sensor window.\\
$\Omega_i$ is calculated using a set of formulas given in reference~\cite{Mathar}. Figure~\ref{trksegment} illustrates track segments and their correspondent $\Omega_i$.\\
The total number of photons arriving (PH) at the ARAPUCA window is obtained as:
\begin{equation}
PH = \frac{1}{4\pi} A_{\Omega}\frac{dN^{\gamma}}{dl}.
\end{equation} 
The expected light calculated by the formulas and the number of PEs measured were compared, as displayed in Figure~\ref{fig:pepattern} for a few example tracks. There is a reduced set of events for which the expected and detected light follow the same shape pattern. It is expected that the absolute efficiency per ARAPUCA is a constant number smaller than 1. As mentioned previously, the efficiency is defined as the ratio between the number of PEs collected by the ARAPUCA and the number of photons arriving at the ARAPUCA window (PH) for a valid track event. Figure~\ref{fig:pepattern} demonstrates, for a valid track Figure~\ref{fig:goodevent1}, the remarkable discrimination power of the ARAPUCA showing how the distribution of the collected light closely follows that expected for the illumination light, scaled down by a constant smaller than one that is directly related to the efficiency of the device. Also, for a different valid track such as in Figure~\ref{fig:goodevent2}, that constant must remain the same within errors. Furthermore, Figure~\ref{fig:goodevent1} and Figure~\ref{fig:goodevent2} show that the ratios are similar for the 8 ARAPUCAs used in the experiment. Figure~\ref{fig:badevent1} and Figure~\ref{fig:badevent2} also show that the same discrimination power allows filtering events where the shape of the distribution of the light collected by the ARAPUCAs do not follow that of the expected illumination. These tracks are due to MIPs that partially illuminated the ARAPUCAs and were mistakenly catalogued by the hodoscope with the wrong geometry due to hot PMTs and coincidental PMT events.\\
The criteria to separate valid from invalid tracks followed a $\chi^2$ requirement as explained in section~\ref{effana2}.

\subsection{Correlation between collected and expected light.}

\begin{figure}[h]
\centering
\includegraphics[width=0.7\textwidth]{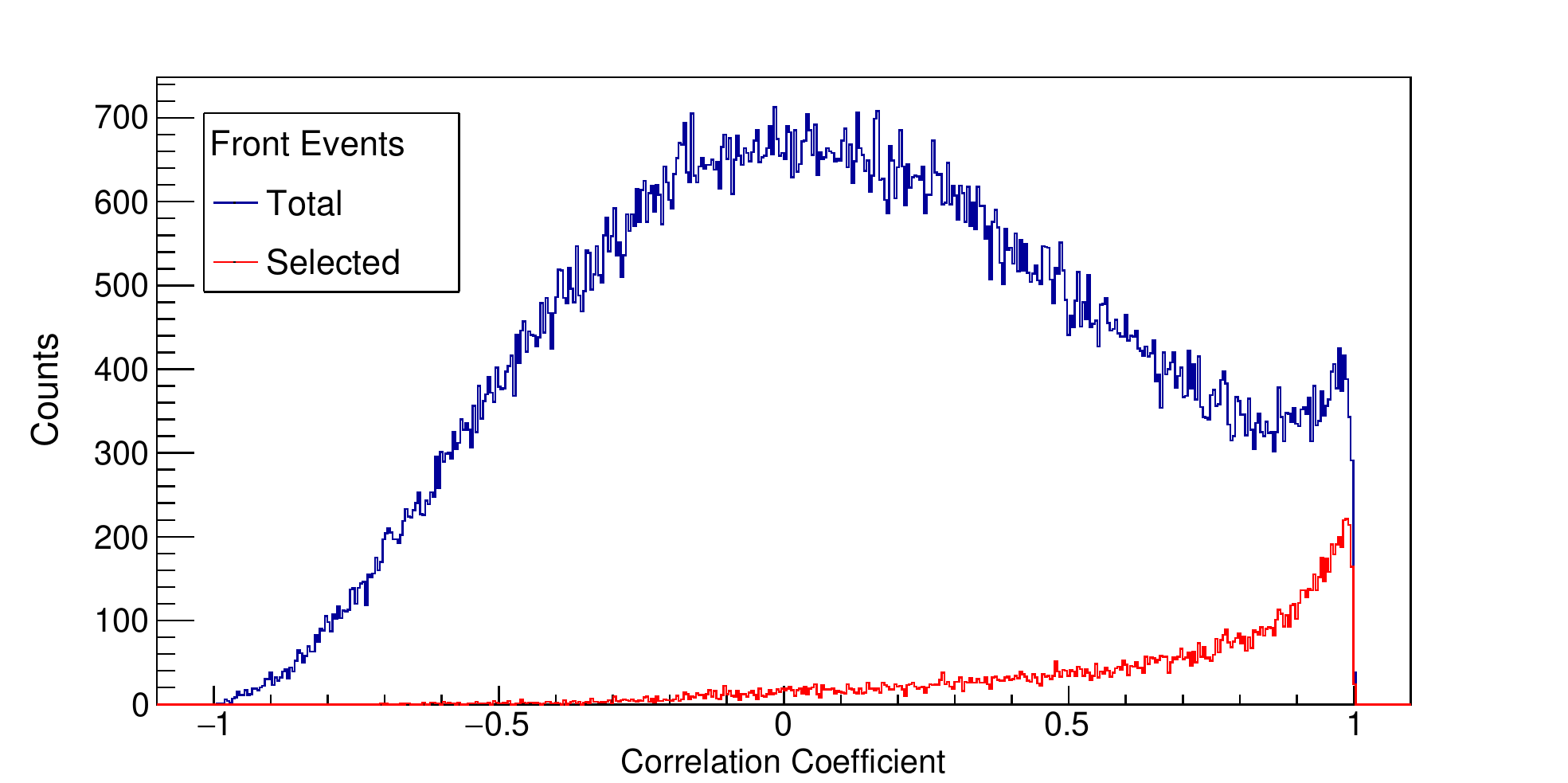}
\caption{Correlation between $PE_i$ and $PH_i$ for the eight ARAPUCAs. Blue is for all the events, red for the events with the PE numbers, recorded by each ARAPUCA, in the range determined from the back and front spectra analysis.}\label{fig:corr}
\end{figure}

After the selection based on front versus back spectrum discrimination, made in section~\ref{bgr}, the events left are 14,811. The cut applied is crude and the data set is still contaminated with background. A second background rejection can be made utilizing the segmentation of the detector given by the 8 ARAPUCA cells shown in section~\ref{lp}.\\
Comparing the amount of expected light coming from the analytical formula and the number of PEs from  waveforms, leads us to the determination of good tracks. In Figure \ref{fig:pepattern} the number of PEs measured (black), the number of landing photons (PH, in red), and their ratio (blue) are displayed for each ARAPUCA, for a few tracks. There is a reduced set of events for which the expected and detected light follow the same pattern Figure \ref{fig:goodevent1} and Figure \ref{fig:goodevent2}. As an estimator for light pattern information we used the correlation between the measured PE number and the calculated number of photons landing on the ARAPUCA for each ARAPUCA in the event. The measure of correlation used is given by the Pearson correlation coefficient defined as
\begin{equation}
C=\frac{\sum \left( PE_i\cdot PH_i \right) -  \sum PE_i \cdot \sum PH_i /8}{\sqrt{\sum (PE_i^2)- \left( \sum PE_i\right)^2/8 } \cdot \sqrt{\sum (PH_i^2)- \left( \sum PH_i\right)^2/8}}
\end{equation}
where $PE_i$ is the number of collected photo electrons in the $i-th$ ARAPUCA and $PH_i$ is the number of expected photons arriving on the $i-th$ ARAPUCA surface. The $i$ index go from 1 to 8.\\
Figure \ref{fig:corr} shows the correlation parameter for the total number of events (blue) and for the events passing the cut on the spectra (red). Most of the events passing the cut are peaked around $C=1$ indicating a good correlation between the eight ARAPUCAs. However there are some events with $C \le 0$. A second requirement on the correlation parameter is done, selecting events with $C\ge 0.7$. After that the total number of selected events is 13005. This number is the total for 366 hours of running time at a rate of $0.010\, Hz$.\\
The number of events selected is in a very good agreement with the rate expected from cosmic ray muon flux through the hodoscopes, obtained analytically, calculating the cosmic ray flux~\cite{angdis} through the hodoscope geometry getting $0.011\, Hz$ and doing a simulation of the muons crossing the hodoscope: $0.0105\, Hz$.\\ 
Figure~\ref{fig:ef1c} shows the common logarithm of the ratio between the sum of the photo electrons measured and the sum of the estimated number of arriving photons on each ARAPUCA, comparing all the data (blue spectra) and the selected dataset (red spectra) using the combination of both spectra selections and correlation requirement.

\subsection{Efficiency analysis}\label{effana}

\begin{figure}
\begin{subfigure}[b]{0.5\textwidth}
\includegraphics[width=\textwidth]{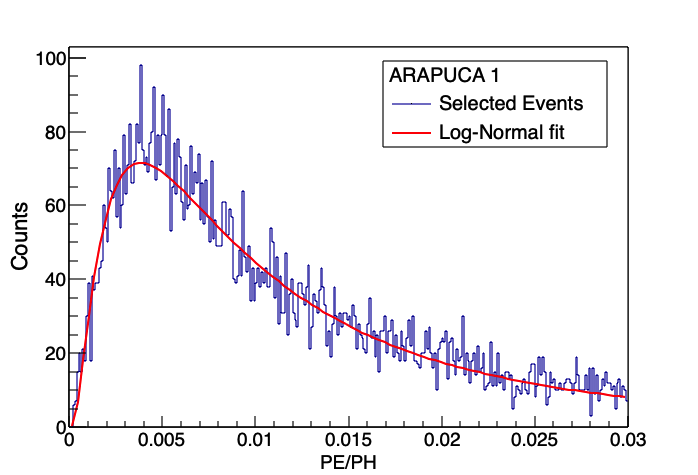}
\end{subfigure}
\begin{subfigure}[b]{0.5\textwidth}
\includegraphics[width=\textwidth]{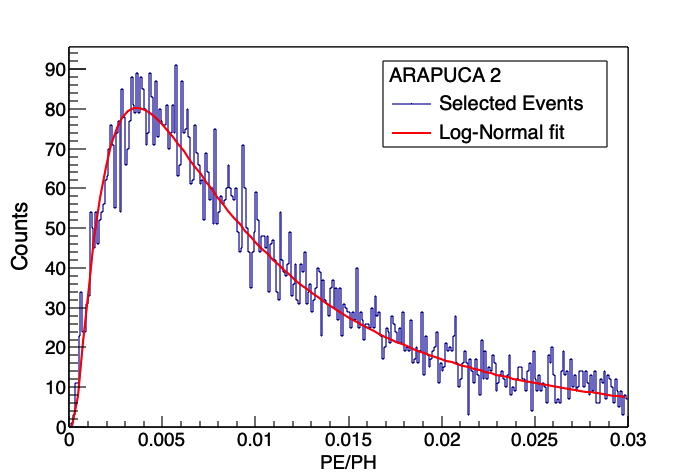}
\end{subfigure}
\begin{subfigure}[b]{0.5\textwidth}
\includegraphics[width=\textwidth]{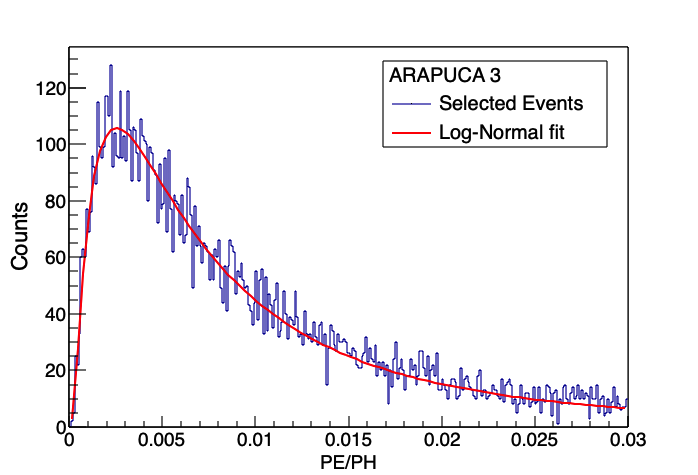}
\end{subfigure}
\begin{subfigure}[b]{0.5\textwidth}
\includegraphics[width=\textwidth]{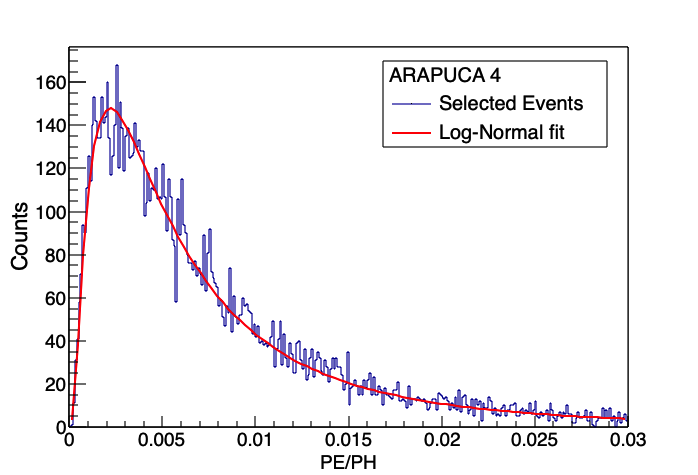}
\end{subfigure}
\begin{subfigure}[b]{0.5\textwidth}
\includegraphics[width=\textwidth]{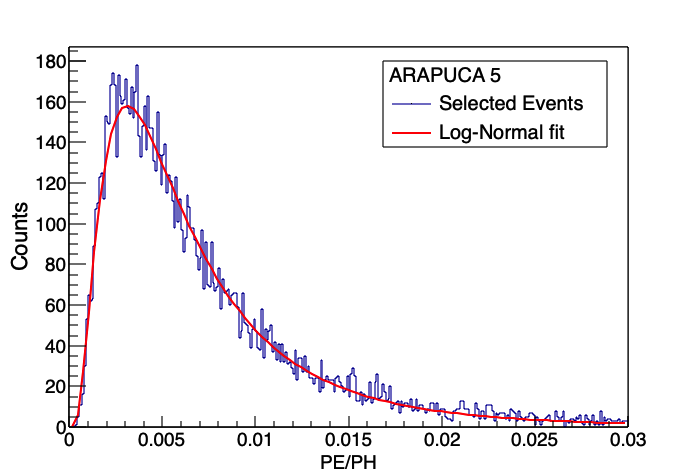}
\end{subfigure}
\begin{subfigure}[b]{0.5\textwidth}
\includegraphics[width=\textwidth]{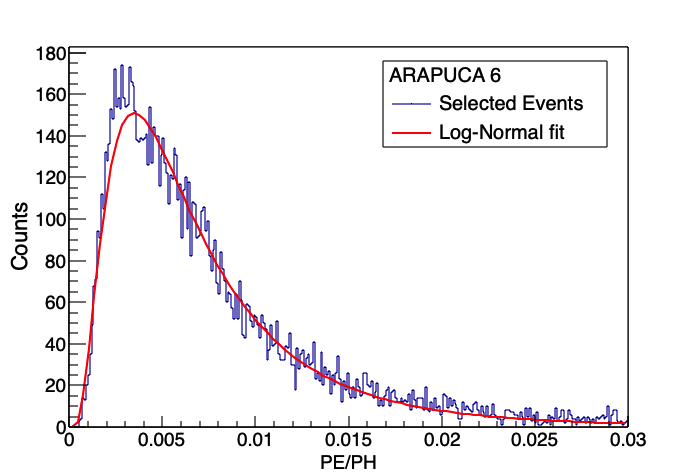}
\end{subfigure}
\begin{subfigure}[b]{0.5\textwidth}
\includegraphics[width=\textwidth]{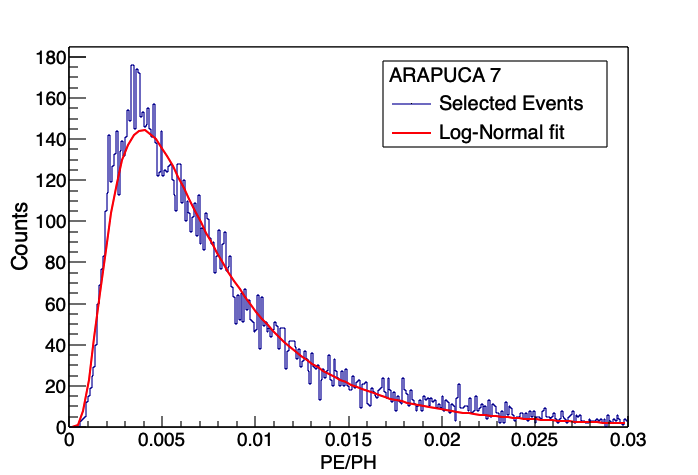}
\end{subfigure}
\begin{subfigure}[b]{0.5\textwidth}
\includegraphics[width=\textwidth]{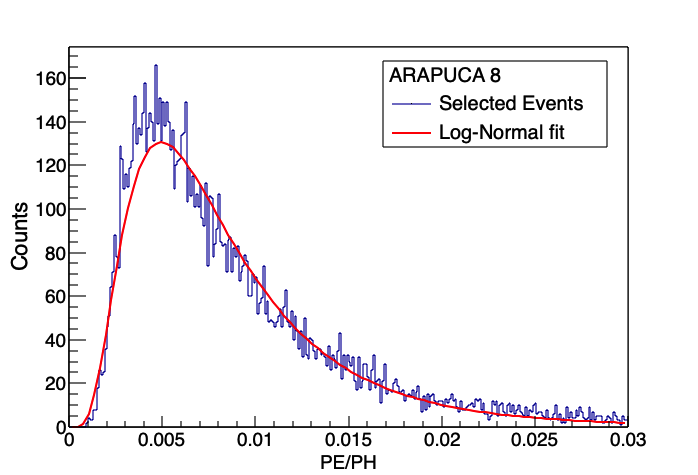}
\end{subfigure}
\caption{Ratio between the photo electrons measured and the estimated number of arriving photons per each ARAPUCA, of the data selected ("$Dataset$ 1A").}
\label{fig:effplots}
\end{figure}

\begin{figure}[h]
\begin{subfigure}[b]{0.5\textwidth}
\captionsetup{width=.9\textwidth}
\includegraphics[width=\textwidth]{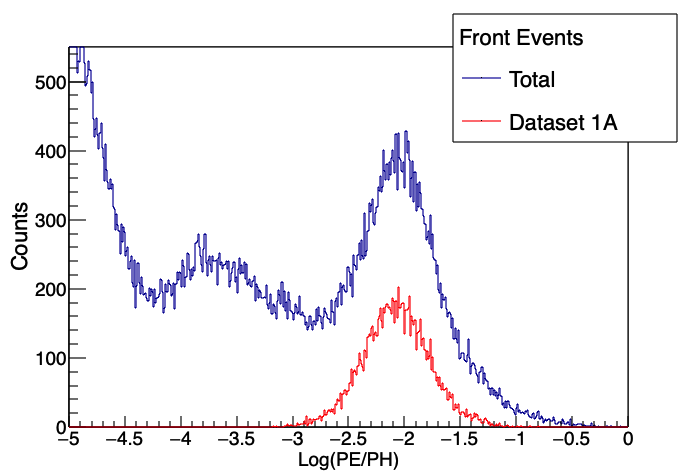}
\caption{Original data set in blue and after the spectra selection and correlation cut in red.}
\label{fig:ef1c}
\end{subfigure}
\begin{subfigure}[b]{0.5\textwidth}
\captionsetup{width=.9\textwidth}
\includegraphics[width=\textwidth]{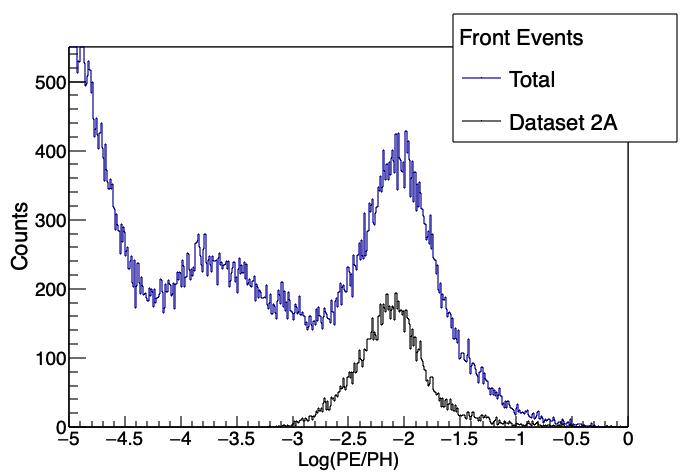}
\caption{Original data set in blue and after a $\chi^2$ selection cut in black using $\Sigma<0.1$ of Eq.~\eqref{tab:eqri3}.}
\label{fig:ef2c}
\end{subfigure}
\begin{subfigure}[b]{0.5\textwidth}
\captionsetup{width=.9\textwidth}
\includegraphics[width=\textwidth]{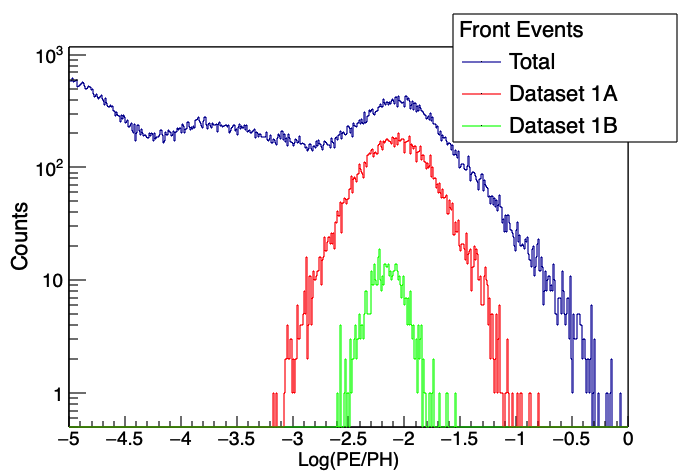}
\caption{Original data set in blue, after the spectra selection and correlation cut in red, and using a further restriction in green.}
\label{fig:ef3c}
\end{subfigure}
\begin{subfigure}[b]{0.5\textwidth}
\captionsetup{width=.9\textwidth}
\includegraphics[width=\textwidth]{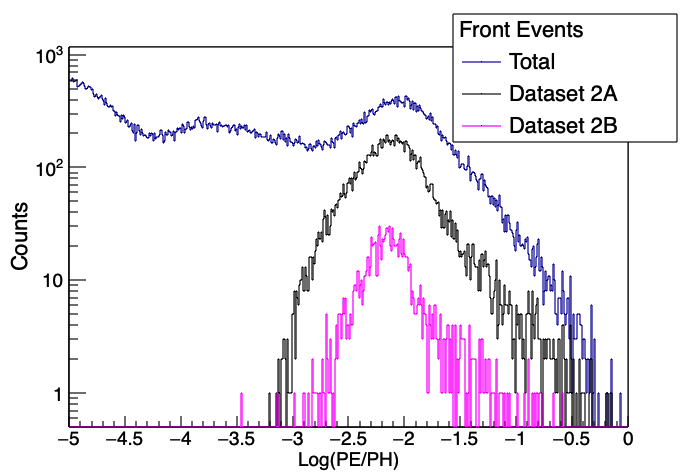}
\caption{Original data set in blue and after a $\chi^2$ selection cut of Eq.~\eqref{tab:eqri3}, black using $\Sigma<0.1$ and purple using $\Sigma<0.02$.}
\label{fig:ef4c}
\end{subfigure}
\caption{Comparison between the two kinds of cut applied to the original data set (blue spectra) of the ratio between the photo electrons collected sum and the estimated number of arriving photons sum. In Figure (a)  and (c) are reported the selected datasets using the cut based on the spectra analysis and correlation condition, in figure (b) and (d) are reported the selected datasets using the $\chi^2$ approach, described in section~\ref{effana} Eq.~\eqref{tab:eqri3}.}
\label{fig:efc}
\end{figure}

The selected data set is then used to perform the efficiency analysis. The efficiency is defined as the ratio between the number of measured photons and the estimated number of arriving photons for each ARAPUCA:
\begin{equation}
\mathcal{R}_{i}=\frac{PE_i}{PH_i}  \label{tab:eqri}
\end{equation} 
Figure~\ref{fig:effplots} shows the ratio distribution for the eight channels. Furthermore, the ratio between the sum of the number of photoelectrons collected by all ARAPUCAs divided by the sum of the number of expected photons landing on all ARAPUCAs is considered.
\begin{equation}
\mathcal{R}_{TOT}=\frac{\sum_{i=1}^8PE_i}{\sum_{i=1}^8PH_i} 
 \label{tab:eqri2}
\end{equation} 
In Figure~\ref{fig:ef1c} is reported $Log(\mathcal{R}_{TOT})$ (for better visualization) for all the data (blue) and the data selected through spectra analysis and correlation cut (red), called "$Dataset$ 1A".\\
Three metrics for analysis were used, finding similar results reported in table \ref{tab:vv}:
\begin{itemize}
\item log-normal fit 
\item robust statistic
\item bootstrap procedure on a reduced dataset
\end{itemize}
The frist two methods were used on the data selected "$Dataset$ 1A" and the last one applying a further restriction on the data, getting a reduced dataset (called "$Dataset$ 1B"). Details about the performance of each metric can be found in the Appendix. Table~\ref{tab:vv} displays the ARAPUCA efficiency values measured by all metrics.\\
\begin{table}[h]
\centering
\begin{tabular}{|c|c|c|c|c|c|c|c|c|} \hline

ARAPUCA  & Mean ($\%$) & Median ($\%$) & Median ($\%$) & MAD ($\%$) \\ 
 & & Fit result & Robust stat. &  \\ \hline \hline

TOT&  0.78 $\pm$ 0.02 &  0.80 & 0.77& 0.41  \\ \hline 
1  &  0.74 $\pm$ 0.02 &  0.80 & 0.75& 0.46 \\ \hline 
2  &  0.77 $\pm$ 0.02 &  0.85 & 0.80& 0.55   \\ \hline
3  &  0.80 $\pm$ 0.02 &  0.84 & 0.77& 0.55   \\ \hline
4  &  0.77 $\pm$ 0.02 &  0.71 & 0.66& 0.42  \\ \hline 
5  &  0.75 $\pm$ 0.02 &  0.67 & 0.64& 0.35  \\ \hline 
6  &  0.77 $\pm$ 0.02 &  0.69 & 0.65& 0.36  \\ \hline 
7  &  0.77 $\pm$ 0.02 &  0.70 & 0.67& 0.37 \\ \hline 
8  &  0.80 $\pm$ 0.02 &  0.80 & 0.75& 0.40  \\ \hline

\end{tabular}
\caption{\label{tab:vv}Ratio values from the three analysis. The error on the fit parameter associated to the median is negligible.}
\end{table}

\subsubsection{$\chi^2$ requirement on light patterns}\label{effana2}

A second criteria of data selection based on a $\chi ^2$ requirement on the ratio $\mathcal{R}_i$, was applied to the original dataset before any cuts (all the events), in order to check the goodness of the spectra selection and correlation requirements.
\begin{equation}
\Sigma=\frac{1}{8}\sum_{i=1}^8\left( \frac{\mathcal{R}_i-\mathcal{R}_{TOT}}{\mathcal{R}_i+\mathcal{R}_{TOT}} \right)^2
\label{tab:eqri3}
\end{equation}
Requiring a condition on $\Sigma$ we can select a certain amount of events. A smaller $\Sigma$ means a better match between collected and expected photons. The conditions on $\Sigma$ were determined in order to get the same amount of data of the other cuts, and the same three analysis were performed: robust statistic and log normal fit for a dataset using $\Sigma<0.1$ ("$Dataset$ 2A") and bootstrap procedure for a reduced dataset using $\Sigma<0.02$ ("$Dataset$ 2B").

\subsubsection{Results}\label{eficana}

The data sets got through the two selection methods were tested incresing the requirements on the cuts, in the first case using the condition shown in Appendix~\ref{strcc}, in the second case using $\Sigma<0.02$ from Eq.~\eqref{tab:eqri3}. These procedures reduced to $\sim 10\%$ both the selected data sets (shown in Figure~\ref{fig:efc}). In both cases the mean values of each ARAPUCA ratio became compatible with the median values found.\\

A similar analysis was made for the Low-Low hodoscope configuration using 3 MeV/cm of energy loss, reflecting the fact that the average muon energy increases for more horizontal events~\cite{energyloss}. Due to the geometry of the tracks only the lower four ARAPUCAs are taken into account in the Low-Low configuration. The upper four have a small acceptance and the ratio $\mathcal{R}_{i}$ (Eq.~\ref{tab:eqri}) is affected by large fluctuations because the number of photons detected is dominated by Poisson statistics. Table~\ref{tab:rrf} shows the results from the two selection criteria for the two configurations. The ARAPUCA efficiency given by the median of the individual cells is $\mathcal{R}=(0.77 \pm 0.02)\%$. The TallBo7 test did not provide a measurement of after pulse and cross talk. The efficiency values reported here do not take into account their contribution. However, an estimation of $(31\pm7)\%$ of after pulse and cross talk is obtained from ProtoDUNE~\cite{protoDUNEpreliminary} preliminary analysis, where the same kind of SiPM were used and run in similar bias conditions. Adjusting by $31\%$ the detection efficiency of the ARAPUCA detectors becomes $\epsilon =(0.60 \pm 0.02)\%$.
\begin{table}
\centering
\begin{tabular}{|c|c|c|c|c|} \hline

ARAPUCA  & Hi-Low & Low-Low  & Hi-Low & Low-Low   \\ 
 & $Dataset$ 1B & $Dataset$ 1B  & $Dataset$ 2B & $Dataset$ 2B \\ \hline \hline

TOT&  0.78 $\pm$ 0.02 &  0.78 $\pm$ 0.02  &0.75 $\pm$ 0.01 &0.75 $\pm$ 0.03 \\ \hline 
1  &  0.74 $\pm$ 0.02 &  - &0.72 $\pm$ 0.01 &- \\    \hline 
2  &  0.77 $\pm$ 0.02 &  - &0.75 $\pm$ 0.01 &-  \\     \hline
3  &  0.80 $\pm$ 0.02 &  - &0.77 $\pm$ 0.01 &-  \\     \hline
4  &  0.77 $\pm$ 0.02 &  - &0.74 $\pm$ 0.01 &-  \\ \hline 
5  &  0.75 $\pm$ 0.02 &  0.81 $\pm$ 0.02  &0.73 $\pm$ 0.01 &0.75 $\pm$ 0.04  \\   \hline 
6  &  0.77 $\pm$ 0.02 &  0.75 $\pm$ 0.02  &0.75 $\pm$ 0.01 &0.78 $\pm$ 0.04  \\    \hline 
7  &  0.77 $\pm$ 0.02 &  0.72 $\pm$ 0.02  &0.76 $\pm$ 0.01 &0.78 $\pm$ 0.04  \\   \hline 
8  &  0.80 $\pm$ 0.02 &  0.79 $\pm$ 0.02  &0.80 $\pm$ 0.01 &0.79 $\pm$ 0.04  \\   \hline

\end{tabular}
\caption{\label{tab:rrf} Mean values of the ratio $\mathcal{R}_{TOT}$ (Eq. \ref{tab:eqri2}) and $\mathcal{R}_{i}$ (Eq. \ref{tab:eqri}) for all the cells. Values for the two hodosocope configurations (Hi-Low ad Low-Low) are reported, using the two selection criteria: spectra analysis and cut on the $\chi^2$ (Eq. \ref{tab:eqri3}).
}
\end{table}

\section{Conclusions}
The Fall 2017 TallBo experiment has been successful in testing ARAPUCA photon detectors. Two of the features of the current ARAPUCA photon detector design were used to be able to filter the data offline. 
The single face detection and the fine segmentation of the ARAPUCAs allowed the offline analysis to separate events from background. The segmentation of the ARAPUCA detector validated the geometry given by the hodoscope. In the end an accurate absolute efficiency of 0.6\% was determined by three different methods. That absolute efficiency is higher than efficiency measurements reported by previous experiments at TallBo using scintillation bars and wavelength shifters.\\
It is also worth mentioning that the ARAPUCAs only used 4 SiPMs per board, the filter to sensor aspect ratio was 35 which shows a remarkable improvement in the equivalent photon collection area given by the light-trap effect in the ARAPUCA.\\
Future work will study improvements in the internal reflective surfaces of the ARAPUCA, wavelength shifter thickness and adherence which should help increasing the light collection. A step forward in the development of ARAPUCA is the active ganging of SiPMs to lower the number of readout channels per module and the use of the two faces of the detector for photon detection, having filters on both sides.\\

\appendix
\section{Analysis metrics}

\subsection{Log-normal fit}
A first analysis is made by fitting the ratio distribution with a Log-Normal distribution:
\begin{equation}
f(x)=\frac{1}{x\sigma\sqrt{2\pi}} e^{-\frac{\left( \ln(x) -\mu\right)^2}{2\sigma^2}} 
\end{equation} 
This choice is driven by the Log$\left(\mathcal{R}_{TOT}\right)$ which seems to follow a Gaussian distribution.\\
The values of the Mean, Median, Mode, Variance and Standard Deviation are obtained by the relations satisfied by the parameters of the Log-Normal distribution:
\begin{itemize}
\item Mean: $e^{\left( \mu + \frac{\sigma^2}{2}\right)}$
\item Median: $e^{ \mu }$
\item Mode: $e^{ \left( \mu-\sigma^2 \right) }$
\item Variance: $\left(e^{\sigma^2}-1 \right)e^{\left( 2\mu + \sigma^2\right)} $
\item Standard Deviation: $ \sqrt{\left(e^{\sigma^2}-1 \right)}e^{\left( \mu + \frac{\sigma^2}{2}\right)} $
\end{itemize} 
The error propagation, from the error in the fit, gives for the median the value:
\begin{equation}
\Delta_{Median}= e^{\mu}\cdot \Delta_{mu}
\end{equation} 

\subsection{Robust statistic}
Because of the presence of outlier events, such as the ones in the far tail of the spectrum, a robust statistic data analysis is made for the data relative to the single ARAPUCA $\left(\mathcal{R}_{i}\right)$ and their sum $\left(\mathcal{R}_{TOT}\right)$, using median and the median absolute deviation (MAD) to build a robust score defined as:
\begin{equation}
\mathcal{S}_i=\frac{\left( \mathcal{R}_i - median_{j=1,...,n}(\mathcal{R}_j) \right)}{MAD}
\end{equation} 
Then the median of the data set composed by the data which passes the requirement $\mathcal{S}_i <2.5$ is calculated.\\
 
\subsection{Strong correlation}\label{strcc}
Finally a restricted data set, obtained with a further strong condition on the light pattern, is analyzed.\\
The strong condition consists in requiring that all ARAPUCAs have the ratio $\mathcal{R}_{i}$ similar to each other:
\begin{equation}
\frac{1}{2} \cdot \mathcal{R}_{j}\le\mathcal{R}_{i}\le 2\cdot \mathcal{R}_{j}
\end{equation} 
This condition reduces the data set to be $\sim \, 10\%$ of the one used in the previous two analysis.\\
A bootstrap procedure is used to find the average values of the ratio and their errors, generating 10 thousand  data sets, each of them composed by random extraction of events, from the original data set.

\section{Acknowledgements}
The authors would like to thank A. Hahn, R. P. Davis, W. Miner, K. Harding for their technical support at PAB. We also thank the whole Indiana University team with whom this experiment was performed. We also thank Eileen Hahn for her invaluable knowledge and support for the wavelength shifter coatings of filters and reflectors, and also Kenneth Treptow for his technical support in assembling the ARAPUCA's module. This work has been partially supported by the Brazilian agency FAPESP under grant no 2017/13942-5. Fermilab is Operated by Fermi Research Alliance, LLC under Contract No. De-AC02-07CH11359 with the United States Department of Energy.

\end{document}